\documentclass[letterpaper]{jpconf}
\usepackage{graphicx}

\newcommand{\msun}{{\mathrm{M}}_\odot}
\newcommand{\Mpc}{\ {\mathrm{Mpc}}}
\begin{document}

\title{LIGO-Virgo searches for gravitational waves from coalescing binaries: a status update}
\author{Anand S. Sengupta for the LIGO Scientific Collaboration and the Virgo Collaboration}
\address{LIGO Laboratory, California Institute of Technology, Pasadena CA 91125, USA.}
\ead{sengupta@ligo.caltech.edu}

\begin{abstract}
Coalescing compact binaries of neutron stars and/or black holes are considered as one of the most promising sources for Earth based gravitational wave detectors. The LIGO-Virgo joint collaboration's Compact Binary Coalescence (CBC) group is searching for gravitational waves emitted by these astrophysical systems by matched filtering the data against theoretically modeled template waveforms. A variety of waveform template families are employed  depending on the mass range probed by the search and the stage of the inspiral phase targeted: restricted post-Newtonian for systems having total mass less than $35 \msun$, numerical relativity inspired complete inspiral-merger-ringdown waveforms for more massive systems up to $100\msun$ and ringdown templates for modeling perturbed black holes up to $500\msun$. We give a status update on CBC group's current efforts and upcoming plans in detecting signatures of astrophysical gravitational waves.
\end{abstract}

\section{Introduction}
Gravitational waves emitted by compact binaries of neutron stars and/or black holes in the last few minutes before merger lie within the sensitive bandwidth of present ground-based interferometric detectors like LIGO \cite{Abbott:2007kva} and Virgo \cite{Acernese:2006bj} and are considered likely candidates for first detection in these instruments. This is partly motivated by the fact that the phase and amplitude evolution of waveforms from such sources are well modeled analytically in the inspiral phase and ringdown phase when the excited final black hole emits gravitational waves and loses energy 
\cite{abbott-2009-80, buonanno:084043}. This allows the construction of templates for use in the application of matched-filtering techniques \cite{SathyaDhurandhar:1991} to filter the noisy data. The intermediate merger phase is not well modeled but can be obtained by numerical relativity \cite{PhysRevLett.95.121101}. These have, in turn,  inspired a whole generation of template waveforms which bridge the inspiral, merger and ringdown parts by various approximations \cite{2009arXiv0909.2867A, buonanno-2007-76}.

The three LIGO detectors at Livingston and Hanford operated at their design sensitivity in a continuous data acquisition mode between November 2005 and September 2007 in their fifth science run (S5), collecting a full year of triple coincident data. The first run of the Virgo detectors (VSR1) coincided with the last 6 months of S5 at a sensitivity comparable to the LIGO instruments above 400 Hz. The aim of this article is to provide a snap-shot of the current activities of the CBC group and present our plans for the near future.

\section{Data processing}
At the core of the CBC group's data analysis pipeline \cite{Allen:2005fk} is the matched filtering engine that correlates data from detectors against theoretical template waveforms expected from the sources. A discrete lattice of templates spanning the intrinsic parameter space of the binary is constructed  such that the maximum loss in signal-to-noise ratio (SNR) is no more than a prescribed limit. Filtering the data $x$ against a template $h(t; \mu_i)$ yields the SNR $\rho$ given by
\begin{equation}
\rho = \frac{\langle x, h \rangle }{\sqrt{\langle h, h \rangle}},
\end{equation}
where  ${\langle x, h \rangle }$ denotes the noise-weighted scalar product of the data and the template. For each template, triggers that have SNR greater than a prescribed threshold are retained. The templates are typically parametrized by the component masses of the coalescing binary (or, for ringdowns, the mass and spin of the final blackhole). These triggers are tagged by the template parameters and coalescence time, which can be later used for clustering \cite{SenguptaTrigScan:2008} and determining coincidence \cite{Robinson:2008}. Although the trigger generation by matched filtering is at the heart of the CBC pipelines, other important steps including: determining detection efficiency and accidental background estimation; vetoing glitchy times; determining coincidence across different instruments and following up interesting coincident triggers; all form important parts of the full data analysis pipeline and are used both for GW-triggered and externally triggered (eg, GRB events) searches.

\section{Status of current activities of the CBC group}
The template families are chosen to optimize the sensitivity of the searches to probe the part of the signal parameter space accessible by the ground-based detectors. The current activities of the group include: the analysis of coincident LIGO S5 and Virgo VSR1 data for low mass coalescing binaries (systems with total mass between $2$ and $35\msun$); analysis of the S5 data for high-mass binaries (systems with total mass $25-100 \msun$) using inspiral-merger-ringdown (IMR) waveforms; those for ringdown of perturbed black holes using exponentially damped sinusoidal waveforms; and analysis of S5 data around times triggered by short, hard gamma-ray bursts. 

\begin{itemize}
  \item[(a)]  The `S5 low-mass search' covers the $2-35\msun$ region of the total-mass parameter space employing frequency domain restricted post-Newtonian (PN) template waveforms accurate to 2PN order in phase evolution up to the innermost stable circular orbit (ISCO). Data is cross-correlated against such template banks consisting of 5-7 thousand templates to provide adequate coverage via semi-automated data analysis pipeline. Historically, it was divided into three separate searches to analyze different subsets of the S5 data: (a) The S5 first calendar year search \cite{Collaboration:2009tt} - sensitive to CBC as far as $150 ~\Mpc$ for systems with total mass $\sim 25\msun$ at an SNR of 8. No gravitational waves were detected by this search. Improved upper limits on the rate of coalescence for neutron star and black hole binary systems were established. (b) The S5 12-18 month analysis covering 186 days beyond the first calendar year \cite{Abbott:2009qj}. No gravitational wave candidate events were reported in this search which yielded a $90\%$-confidence rate upper limit of CBC of neutron stars at $1.4\times 10^{-2}$ per year per $L_{10}$ where the latter is $10^{10}$ times the blue solar luminosity. (c) The last six months of S5 data was analyzed in coincidence with VSR1 allowing the first joint LIGO-Virgo data analysis. This work is nearing completion and publication. The upper limits are cumulative (i.e, they include the results from all previous analyses) and are combined in a Bayesian manner.  

\begin{figure} 
\begin{center} 
\includegraphics[width=0.85\textwidth]{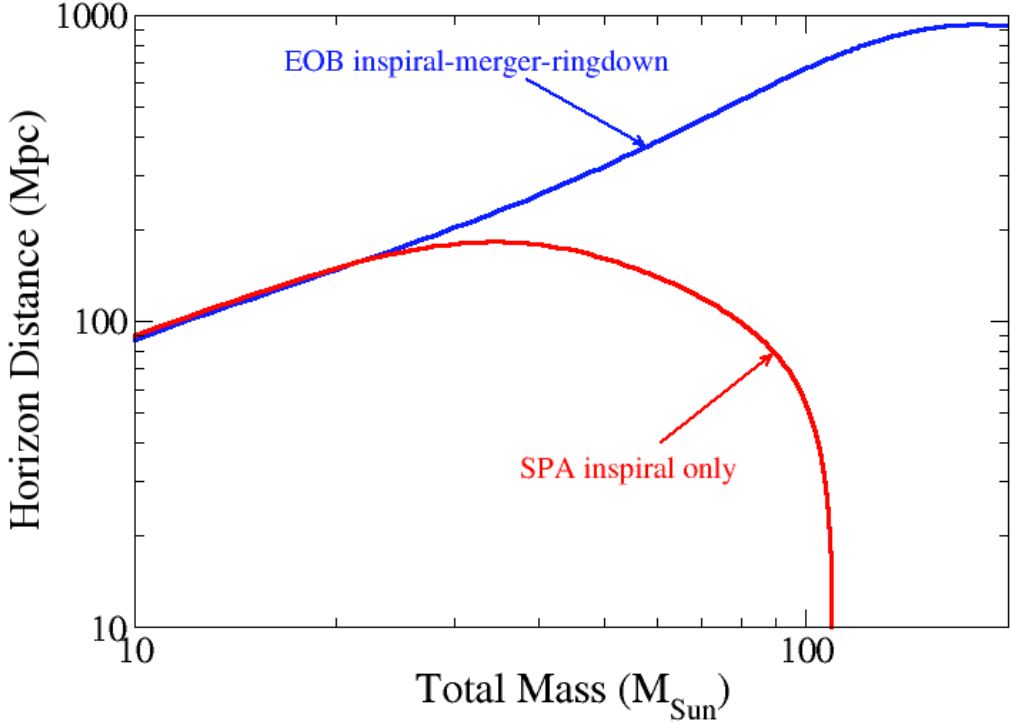} 
\end{center} 
\caption{Horizon distance (distance at which an optimally located and oriented binary would be detected in the LIGO detectors with SNR of 8) for restricted 2PN (red trace) and EOBNR inspiral-merger-ringdown waveforms tuned to numerical relativity; for equal-mass binaries where the symmetric mass ratio $\eta = 1/4$. Initial LIGO design noise power spectral density is assumed. To convert the horizon distance for other mass ratio, we need to multiply by $\sqrt{4\eta} $. Note that the high-mass search using time-domain EOBNR waveforms greatly extends the reach of the search for systems with total mass above $\sim 25 \msun$. } 
\end{figure} 
  
  \item[(b)] The `S5 high-mass search' uses time-domain numerical-relativity tuned effective one-body (EOBNR) \cite{buonanno-2007-76} template waveforms to cover the total mass space between $25-100\msun$ with some overlap with the low mass search. The frequency domain templates used in the low mass search are not suitable in this case as a significant portion of the SNR comes from the merger and ringdown phase of the coalescence for these massive systems and are not captured by the post-Newtonian template waveforms. The high-mass search compliments the low-mass search by extending the boundaries of the parameter space explored and also the volume reach of the search. At present, spin effects are not taken into account in the template bank used for this search. We hope to take spin effects into account by running spinning waveforms through to evaluate efficiencies. Filtering, coincidence and signal-based veto techniques used in this search are similar to the low-mass search as is the technique of `effective SNR' ranking of the coincident triggers. The S5 data is being analysed on a month by month basis. New structured query language (SQL)-based post-processing tools developed are being tried out on this search by the CBC group for the first time.
  
    \item[(c)] Recently, the CBC group completed the first ringdown search \cite{abbott-2009-80} using the fourth LIGO science data (S4) spanning black holes with masses between $10-500\msun$ with spins ranging from $0 - 0.994$. The waveform templates are modeled as damped sinusoids spanning this two-dimensional parameter space.  It is currently gearing up to analyze the S5 data with an eye to determining coincidence with the triggers from the S5 high-mass search. Although the fundamental data analysis pipeline as outlined in S4 paper has been retained for S5, a new coincidence algorithm has now been deployed which lowers the false alarm rate and effectively makes the search more sensitive. Tuning of the search pipeline is in progress using complete IMR and ringdown-only waveforms.

  \item[(d)] Coalescence of two neutron stars or a neutron star and black hole are considered as likely  progenitors of short gamma-ray bursts (GRBs). The CBC group has analyzed the science data around 22 such GRBs that occurred over the duration of S5/VSR1 to look for gravitational-wave counterparts, employing restricted 2PN templates with total mass between $2-35\msun$. Although the actual number of short hard GRBs was larger, only 22 had enough science data available across multiple instruments to guarantee at least a two-interferometer coincidence. In this search both time and directional coincidence for triggers are required, enabling a lowering of SNR thresholds to enhance the sensitivity of the search. An earlier analysis of GRB 070201 \cite{collaboration-2007} (whose sky position error box overlaps the M31 Andromeda galaxy) was repeated in view of the subsequent improvement to the data analysis pipeline. The earlier analysis carried out by the CBC group lead to a null result excluding a CBC progenitor in M31 ( at $99\%$ confidence).  
\end{itemize}

Apart from these searches, the CBC group is also developing techniques to follow-up interesting candidate events with a view to developing detection confidence. These include using Markov-chain Monte-Carlo methods \cite{vandersluys-2009-204010} to estimate the parameters of detection candidates. Other activities include: developing better IMR waveforms that are faithful to the numerical waveforms over a wider range of parameters; including spin as a parameter in our detection templates - especially for high-mass end of the parameter space; detector characterization to understand the source of coincident glitches; and preparing the ground for upcoming, more sensitive science runs. 

\section{New data analysis techniques on the horizon}
Since the beginning of S5, the basic data analysis pipeline has seen several improvements to its core functionality and many new features have been added. In this section, we highlight some of the most important ones. A key area of improvement has been the addition of numerical relativity inspired waveforms to the LSC and Virgo collaboration's arsenal of existing templates: time-domain EOB waveforms calculated up to the light ring, calibrated to numerical simulations and then matched to the quasi-normal modes of the ringdown at the end, have extended the effective bandwidth of the high-mass search thereby increasing the distance reach and sensitivity of the high-mass search; see Figure 1. Phenomenological IMR waveforms \cite{2009arXiv0909.2867A,2008PhRvD..77j4017A} that include spins (albeit restricted to non-precessing aligned/anti-aligned configurations) now allow us to not only probe significantly larger parts of the CBC  parameter space but also to inject such signals into the data stream for more reliable estimates of the efficiency of our pipeline in detecting gravitational waves from CBCs. Several new analysis technologies are on the horizon: hardware-accelerated signal processing where the computer's graphical processing units (GPU) are used to parallelize the fast-Fourier transform calculations have been tested and found to make our search pipeline an order of magnitude faster. Multi-variate statistical classification tools are being developed to bring in machine-learning algorithms that are able to classify our putative events and better distinguish them from background \cite{LIGO-SURF-Miller}. New ranking statistics based on likelihood ratios have been found to perform better than traditional `effective SNR' statistic and have been deployed in some of the searches. The coherent inspiral pipeline has matured to the point where it can be run in a follow-up mode for interesting candidates that pass the stringent mass and time coincidence tests. Efforts are also on towards a Bayesian analysis of data from arbitrary network of detectors for the extraction of parameters of a likely signal in the data \cite{veitch-2008-25}. The full posterior probability density function for parameter estimation (using a wide range of template waveforms) can be computed, allowing estimation of likely parameters and the range of their uncertainty. These new techniques are at various stages of development and stem from a common desire of the group to work towards building confidence in our first detection.

\section{Future plans: S6, VSR2 and beyond}
Following S5 and VSR1 data taking, the detectors are undergoing scheduled upgrades which will improve the sensitivity in these instruments and the volume of space that these `enhanced' detectors can probe. Data taking with the enhanced detectors (LIGO's S6 and Virgo's VSR2) began in June 2009 and will continue for approximately 1.5 years. At that point, the detectors will undergo more major upgrades. Advanced LIGO is scheduled to be fully commissioned in 2015. In these instruments, the improvement in sensitivity is expected to be a factor of 10 over initial LIGO and a 1000 fold increase in the event rates, making the detection of gravitational waves routine. S6/VSR2 data is being used as a test bed for several data analysis techniques developed by the CBC group in the recent past. A major paradigm shift has been to focus on rapid turn around of the  results by analyzing a week's data at a time. Low latency pipelines being developed by the group have been deployed and are being extensively tested. Many of the post-processing tasks and follow-up of interesting triggers from the main pipeline are being automated: among them, rapid sky localization and parameter estimation; and improved background estimation are being explored. The latter is motivated by a chance for electromagnetic followup by quick identification and localization of a transient corresponding to an interesting gravitational-wave candidates. This is also establishing the protocol for external collaborations with other ground and space based telescopes and the exciting opportunity to `see' a gravitational wave trigger in different wavelengths.

\ack{I would like to thank Szabi Marka, Zzusa Marka and other organisers of the 8th E. Amaldi conference in New York City. I also thank Ajith Parameswaran, Craig Robinson and Alan Weinstein for reading and editing the manuscript, comments and feedback.
}

\section*{References}
\bibliography{ref}

\end{document}